# Tailoring the photon hopping by nearest and next-nearest-neighbour interaction in photonic arrays

**Niccolò Caselli[1,2]\*, Francesco Riboli[3], Federico La China[1,2], Annamaria Gerardino[4], Lianhe Li[5], Edmund H. Linfield[5], Francesco Pagliano[6], Andrea Fiore[6], Francesca Intonti[1,2], and Massimo Gurioli[1,2]**

[1] European Laboratory for Non-linear Spectroscopy, 50019 Sesto Fiorentino (FI), Italy

[2] Department of Physics, University of Florence, 50019 Sesto Fiorentino (FI), Italy

[3] Department of Physics, University of Trento, via Sommarive 14, 38123, Povo (TN), Italy.

[4] Institute of Photonics and Nanotechnology, CNR, via Cineto Romano 42, 00156 Roma, Italy

[5] School of Electronic and Electrical Engineering, University of Leeds, Leeds LS2 9JT, United Kingdom

[6] COBRA Research Institute, Eindhoven University of Technology, 5600 MB Eindhoven, The Netherlands

\*E-mail: caselli@lens.unifi.it

**ABSTRACT:**

Arrays of photonic cavities are relevant structures for developing large-scale photonic integrated circuits and for investigating basic quantum electrodynamics phenomena, due to the photon hopping between interacting nanoresonators.

Here, we investigate, by means of scanning near-field spectroscopy, numerical calculations and an analytical model, the role of different neighboring interactions that give rise to delocalized supermodes in different photonic crystal array configurations. The systems under investigation consist of three nominally identical two-dimensional photonic crystal nanocavities on membrane aligned along the two symmetry axes of the triangular photonic crystal lattice. We find that the nearest and next-nearest-neighbour coupling terms can be of the same relevance. In this case, a non-intuitive picture describes the resonant modes, and the photon hopping between adjacent nano-resonators is strongly affected. Our findings prove that exotic configurations and even post-fabrication engineering of coupled nanoresonators could directly tailor the mode spatial distribution and the group velocity in coupled resonator optical waveguides.

**INTRODUCTION:**

Coupled optical cavities, due to their ability in modifying the light matter interaction, play a relevant role in the study of cavity quantum electrodynamics phenomena and in the development of new photonic-based applications. Electromagnetically induced transparency-like effects and group velocity tailoring have been demonstrated in systems with many coupled resonators [1-5]. These systems also proved to be a reliable platform to test topological invariants with photons [6]. While two coupled cavities, by means of Purcell effect, can realize a bright source of entangled photon pairs as well as an ultra-fast control of the emission rate of embedded quantum emitters [7-8]. In addition, the significant system of three coupled cavities has led to the proposal of an optical analogue of the Josephson interferometer and has been recently proposed for mode tailoring in quantum electrodynamics experiments [9-10]. Strongly interacting resonators would represent the basis for developing two-qubit gates in integrated structures [11]. Moreover, coherent cavity quantum electrodynamics effects have been predicted in photonic arrays formed by many optical cavities exhibiting a non-linearity [12-13].

Within this framework, photonic crystal cavities (PCCs) represent promising building blocks, since they generate a discrete series of localized light states with very high quality factors, up to $10^6$, and very small mode volumes, down to $(\lambda/n)^3$ [14-17]. Therefore, a single PCC can be considered as the optical analogue of a bounded electronic quantum system that is characterized by a discrete set of energy levels with a finite spectral width and spatial extent. Two or more interacting nanocavities give rise to the hybridization of the single photonic orbitals, similarly as the inter-atomic coupling drives the molecular orbital in real molecules. Interacting PCCs are therefore called photonic arrays (or molecules if the number of PCCs is small) [18-20]. Photonic arrays formed by nominally identical cavities, like coupled resonator optical waveguide (CROW), exhibit spatially delocalized optical modes and spectral minibands, which describe the light "hopping" between adjacent resonators [5,21-23]. CROW systems are therefore highly promising elements for developing large-scale photonic integrated circuits, in close analogy with electric circuits [24]. However, extending the analogy between electron states and light states is noteworthy, but not completely straightforward [25-28]. For instance, despite the large interest in the field, very little is known on the role of different neighbour interactions between array of PCCs.

Here, we approach the array of three coupled PCCs in two configuration geometries, leading to different behaviours in terms of mode splitting and spatial distribution, as the inter-cavity alignment changes. We demonstrate that the coupling between PCCs can be driven by the interplay between the nearest-neighbour and the next-nearest-neighbour interactions. When the nearest-neighbour coupling coefficient is the dominant term, the photonic array modes behave very similarly to quantum system, such as coupled quantum wells, concerning the mode envelope distributions [29]. On the contrary, by designing a photonic array of PCCs where the role of the next-nearest-neighbour coupling is not negligible, we obtain a non-intuitive picture for the mode splitting and spatial distributions.

**METHODS:**

We use a 320 nm-thick GaAs membrane with three central layers of self-assembled high-density InAs quantum dots (QDs) emitting at room temperature a broad spectrum centred at about 1270 nm. The two-dimensional photonic crystal built in the membrane plane consists of a triangular lattice of air holes with lattice parameter a=311 nm and air filling fraction equal to 35%. The holes are fabricated by electron-beam lithography and subsequent chemical etching [30]. The single PCC is formed by four missing holes organized in a diamond-like geometry, denominated D2 nanocavity. We investigate the photonic array that consist in three adjacent D2 nanocavities aligned along the M or the K-axis of the photonic crystal reciprocal lattice. A room temperature commercial scanning near-field optical microscope (SNOM) is used in illumination/collection configuration. The sample is excited by a diode laser (780 nm) and the emitted photoluminescence (PL) signal is coupled to a spectrometer and it is finally collected by a liquid nitrogen cooled InGaAs array. This setup gives a combined spectral and spatial resolution of 0.11 nm and 250 nm, respectively. In order to theoretically evaluate the spectral behaviour and the mode distributions of the nominal structure, numerical calculations are performed with a commercial three-dimensional finite-difference time domain (FDTD) code, while the band diagrams numerical calculations are performed by a two-dimensional plane wave expansion algorithm.

The optical properties of the D2 nanocavity have been widely studied [31-32]. For sake of simplicity, in the following analysis we focus the attention only on the coupling between the lower energy mode of each single PCC. This photonic mode is mainly elongated along the M-axis of the photonic crystal reciprocal lattice, as inferred from Fig. 1a)-b). This behaviour accounts for a large mode overlap between adjacent PCCs in the M-aligned array and to a large nearest-neighbour interaction. On the contrary, the alignment along the K-axis corresponds to a small overlap between adjacent PCCs that results in a smaller photonic hopping.

**RESULTS AND DISCUSSION:**

In order to theoretically investigate the differences in the two kinds of alignment we perform two-dimensional plane wave expansion calculations, by modelling an infinite array of PCCs aligned along the M or the K-axis, whose schematics are reported in the insets of Fig. 1c)-d), respectively. We obtain the energy band diagram projected along the wavevector component parallel to the coupling axis for both the configurations, as reported in Fig. 1c)-d). The finite width of both dispersion curves demonstrates the effective photonic hopping between the aligned nanocavities [21].

The expected interaction difference between the two kinds of alignments is also confirmed by modal dispersion analysis, which shows clear minibands for both array configurations but with different trends. In particular, for the M-alignment, a monotonic and large increase of the energy of the photonic guided mode as a function of $k_y$ is found, as reported in Fig. 1c). This pattern corresponds to a photonic guided mode with positive group velocity. On the contrary, in the K-alignment we find a small and non-monotonic dispersion for the photonic mode, as reported in Fig. 1d). This is a signature of anomalous light propagation that shows a change in the sign of the group velocity, from negative to positive, close to $k_xD/2\pi=0.25$. Therefore, in the K-alignment also the interaction between distant neighbours has to be considered. Within the tight binding approximation, it is possible to estimate the amplitude of each neighbour coupling term for both array alignments by fitting the resulting miniband with the formula [22,33]:

$$\hbar\omega(k) = \hbar\omega_0 - \sum_{m=1}^{\infty} 2 J_m \cos(mkD) \qquad (1)$$

where $k$ represents the wavevector component considered, $\hbar\omega_0$ is the energy of the isolated single cavity, $D$ is the spatial separation between adjacent PCCs and the parameters $J_m$ are the nearest-neighbour ($m=1$), the next-nearest-neighbour ($m=2$), and successive neighbouring ($m>2$) coupling terms. They can be expressed as [22,33]:

$$J_m = \frac{\hbar\omega_0 \int \Delta\varepsilon(\vec{r})\vec{E}_0(\vec{r}) \cdot \vec{E}_0(\vec{r} + m\vec{D}) d^3r}{\int \left[ \mu_0 |\vec{H}_0(\vec{r})|^2 + \varepsilon(\vec{r}) |\vec{E}_0(\vec{r})|^2 \right] d^3r} \qquad (2)$$

where the axes origin is considered in the centre of the lower side PCC for M-aligned array and of the left side PCC for K-aligned array, $\vec{E}_0$ and $\vec{H}_0$ represent the electric and magnetic field localized in the isolated single cavity; $\Delta\varepsilon = \varepsilon' - \varepsilon$ is the difference of spatial-dependent dielectric constant between the single PCC ($\varepsilon$) and the photonic array ($\varepsilon'$); $\vec{D}$ is represented by $\vec{D} = 3\sqrt{3}a\hat{y}$ for the M-alignment and by $\vec{D} = 3a\hat{x}$ for the K-alignment.

We fit the calculated energy dispersion with Eq.(1), considering only the first three $J_m$. Note that the wavelength of the guided mode is about 400 nm and the $m=3$ neighbouring term links cavities that are separated by 4.8 µm and 2.8 µm in the M and in the K-aligned array, respectively. We find that for the M-alignment the modal dispersion of the miniband is governed by the coupling between adjacent PCCs ($J_1/a = -6.4$ meV), while the amplitudes of the higher order contributions are negligible (see Table 1). The case of K-aligned array is quite different. The values of $J_1$ and $J_2$ have the same order of magnitude ($J_1/a = 0.58$ meV, $J_2/a = 0.78$ meV), while only $J_3$ is negligible (see Table 1). Therefore, in the K-aligned array the approximation of only nearest-neighbour interaction fails and in any finite array a complex mode splitting and non-intuitive delocalization of the resonant modes is expected.

Moreover, the signs of $J_1$ and $J_2$ in the K-aligned geometry are opposite with respect to the M-aligned case. The negative (positive) coupling strength gives rise to a bonding (antibonding) ground state in M-aligned (K-aligned) arrays of any length, as it has been already investigated for the system of two coupled D2 nanocavities [25]. This is a direct consequence of the interference effects between the oscillating tails of the single localized modes, which can be constructive or destructive, giving rise to $J_m$ of different sign as a function of the inter-cavity distance, as expressed by Eq.(2). On the contrary, the coupling between electrons in adjacent quantum wells has a fixed sign as it is realized by the monotonically evanescent tails of the wavefunctions. Note that, exploiting the interference effect on $J_m$ in photonic crystals, our configuration of the K-aligned arrays approaches the condition of vanishing $J_1$ [28] and its small absolute value is the main reason of $J_1 \sim J_2$.

|  | $J_1$ [$a/\lambda$] | $J_2$ [$a/\lambda$] | $J_3$ [$a/\lambda$] |
| --- | --- | --- | --- |
| M coupling | $-1.6 \cdot 10^{-3}$ | $-4.2 \cdot 10^{-5}$ | $-1.8 \cdot 10^{-5}$ |
| K coupling | $+1.5 \cdot 10^{-4}$ | $+1.9 \cdot 10^{-4}$ | $-3.5 \cdot 10^{-5}$ |

**Table 1**: Values of the coupling terms $J_m$, in $a/\lambda$ units, calculated for the two-dimensional infinite array of D2 nanocavities aligned along the M and the K-direction, respectively. The values are obtained by fitting with Eq.(1) the dispersion relations reported in Fig. 1c)-d). For both alignments $J_3$ is much smaller than $J_1$, therefore it is demonstrated the validity of considering only $m \leq 3$ in Eq.(1).

In order to validate the properties extracted from the modal dispersion analysis, we experimentally investigate the photonic molecule array composed by three D2 nanocavities aligned either along the M or the K-axis.

The SEM image of the investigated M-aligned photonic array of 3 PCCs is reported in the inset of Fig. 2a). The corresponding typical near-field PL spectrum, spatially averaged over the SEM image, is reported in Fig. 2a) and it shows three peaks almost equally spaced by about 10 nm, hereafter denominated $T_n$ (n=1,2,3 for decreasing wavelength). The presence of three resonant modes is in agreement with the theoretical FDTD calculations, whose spectrum is reported in Fig. 2c). The spectral offset between experiment and theory has to be ascribed to the considered slab refractive index, which is evaluated as GaAs refractive index at the wavelength of interest neglecting any possible oxidation or dispersion of the material (n=3.484), and to a combined effect of the inhomogeneity of the membrane thickness and of the fabrication-induced disorder in the photonic crystal pores. Figure 2b) shows the summary of the $T_n$ peak positions (obtained by a Lorentzian function fit of each peak) for different samples, which possess nominal identical design. A spread in the peak positions of the order of few nanometres is clear, but the separations between $T_1$-$T_2$ and $T_2$-$T_3$ have only slight fluctuations. This means that disorder induced detuning and inhomogeneity of the membrane thickness have a limited impact on the mode interaction strength for the M-alignment geometry.

The case of the K-aligned array of 3 PCCs, whose typical SEM image is reported in the inset of Fig. 2d), is more complex. The summary of the measured $T_n$ peak positions are shown in Fig. 2e) for nominally identical structures. The overall splitting ($T_1$-$T_3$) is much smaller with respect to the M-alignment, denoting a weaker mode coupling. Moreover, a large variation of the mode splitting is reported, as shown in Fig. 2e), thus indicating that the disorder induced detuning can be comparable with the mode coupling. The investigated K-aligned structure where the role of disorder is less pronounced, is likely the array#1, whose PL spectrum is reported in Fig. 2d). It shows an almost degeneration of $T_1$ and $T_2$ modes. In fact, we find that array#1 represents the experimental realization that more closely corresponds to a system of nominally identical PCCs, as inferred by the comparison between the experimental data of Fig. 2d) and the FDTD calculated spectrum reported in Fig. 2f).

In order to understand the physics underlying the mode splitting of the two kinds of arrays we develop a minimal model of three coupled resonators. The behaviour of the M-aligned array can be reproduced with the nearest-neighbour coupling only. It is represented by $g_M = -|g_M|$, which is negative and gives rise to a bonding ground state [25]. Assuming three identical resonators with energy $\hbar\omega_0$, thus neglecting any detuning, the energies and distributions of the molecular modes are found by solving the eigenvalue problem of the following 3x3 matrix:

$$M = \begin{pmatrix} \hbar\omega_0 & -|g_M| & 0 \\ -|g_M| & \hbar\omega_0 & -|g_M| \\ 0 & -|g_M| & \hbar\omega_0 \end{pmatrix} \qquad (3)$$

the solutions are the eigenvalues $E_n$ and the eigenvectors $c_n$ given by:

$$E_1 = \hbar\omega_0 - \sqrt{2}|g_M| \qquad c_1 = (1, \sqrt{2}, 1)$$
$$E_2 = \hbar\omega_0 \qquad c_2 = (-1, 0, 1) \qquad (4)$$
$$E_3 = \hbar\omega_0 + \sqrt{2}|g_M| \qquad c_3 = (1, -\sqrt{2}, 1)$$

where the three components of the eigenvectors describe the envelope mode amplitude on the basis of the isolated resonators. The energy splitting between the lower and the higher resonant mode ($2\sqrt{2}|g_M|$) is $\sqrt{2}$ times larger than the splitting of two coupled resonators only. This prediction is in good agreement with the experimental comparison between the average splitting of 20 nm between $T_1$ and $T_3$ found in the investigated M-aligned system and the 13 nm splitting observed in a two PCCs M-aligned system, with identical coupling region [32]. In particular, if the detuning is neglected the average experimental $T_1$ - $T_3$ splitting together with the analytical expression of $|E_3 - E_1|$ provides $|g_M| = 5.2$ meV. In addition, the FDTD calculated $T_1$ - $T_3$ splitting [Fig. 2c)], allows us to retrieve the expected coupling strength for a three-dimensional system on slab, which is equal to $|g_M| = 5.6$ meV, comparable to the experimental result. Moreover, these results are consistent with $J_1/a = -6.4$ meV, as retrieved from the band diagram analysis and reported in Table 1. In order to investigate the sub-wavelength spatial distribution of the photonic array modes we map the PL intensity at the peak wavelength for each resonant mode as a function of the SNOM tip position. The comparison between the PL intensity maps of the $T_n$ modes of the M-aligned array#2 and the corresponding electric field intensity distributions calculated by FDTD is reported in Fig. 3. The $T_1$ and $T_3$ modes are mainly localized in the central PCC, thus corresponding to the $c_1$ and $c_3$ states evaluated from the model of matrix M, see Equations (4); while the $T_2$ mode is mainly localized in the external PCCs, as expected for the $c_2$ state. This comparison also confirms the validity of the tight-binding approximation in reproducing the investigated photonic molecule. In fact, the mode distributions evaluated by FDTD indicate that all the molecular modes are roughly formed by a linear combination of the single cavity modes. Concerning the K-aligned array of PCCs geometry a quite different behaviour is observed. In this configuration the next-nearest-neighbour interaction plays a crucial role, as evidenced by the ratio $J_2/J_1 \approx 1.3$ of Table (1).

The photonic array of three PCCs denoted as K-aligned array#1 is the structure where the disorder induced detuning is mostly negligible. Due to the fact that the three peaks in the PL spectra are characterized by a full width at half maximum comparable to their spectral distances, see Fig. 2d), we performed a three Lorentzian function fit in order to define the intensity contribution of each peak at every tip position. Following this procedure Fig. 4 shows the PL measurements of the mode intensity distributions. In particular, the higher quality factor mode ($T_3$) exhibits an intensity distribution mainly delocalized over the two external nanocavities, as reported in Fig. 4e). This distribution is in agreement with the calculated FDTD electric field intensity map, shown in Fig. 4f). Concurrently, the $T_1$ and $T_2$ PL distributions are delocalized over both central and left side PCCs, as reported in Fig. 4a) and c), respectively. Similar results are retrieved by the FDTD calculated electric field intensity maps of the $T_1$ and $T_2$ modes shown in Fig. 4b) and d). Nevertheless, for a better agreement of the relative intensities, an exchange of the $T_1$ and $T_2$ modes with respect to the experimental case should be considered. This exchange behaviour can be driven by a little amount of detuning (even much smaller than the coupling strength), that slightly shifts $T_1$ with respect to $T_2$ or vice versa.

In order to confirm this description of the K-aligned array of three PCCs, we employ the model of three coupled resonators including also the contribution of the next-nearest-neighbour coupling $d_K = |d_K|$ besides the nearest-neighbour term $g_K = |g_K|$, which this time are both positive. The resonant energies and distributions are found by solving the eigenvalue problem of the following 3x3 matrix:

$$K = \begin{pmatrix} \hbar\omega_0 & g_K & d_K \\ g_K & \hbar\omega_0 & g_K \\ d_K & g_K & \hbar\omega_0 \end{pmatrix} \qquad (5)$$

we compare the difference between the eigenvalues of K-matrix (5) with the mode splitting of the FDTD spectrum reported in Fig. 2f), which are $T_1$-$T_2$ = 0.35 nm and $T_1$-$T_3$ =2.94 nm. We obtain the same splitting of Fig. 2f) by using $g_K$ =0.69 meV and $d_K$ =0.88 meV. Moreover, these results agree with the $J_1/a$ =0.58 meV and $J_2/a$=0.78 meV values retrieved from the modal dispersion analysis of the K-aligned infinite array and exhibit a similar ratio ($d_K/g_K \approx 1.3 \approx J_2/J_1$).

The corresponding eigenvalues, evaluated in wavelength units as $\lambda_n = \hbar c E_n^{-1}$, and eigenvectors are:

$$\lambda_1 = 1306.26 \text{ nm} \qquad c_1 \cong (-0.71, 0, 0.71)$$
$$\lambda_2 = 1305.92 \text{ nm} \qquad c_2 \cong (0.38, -0.84, 0.38) \qquad (6)$$
$$\lambda_3 = 1302.98 \text{ nm} \qquad c_3 \cong (0.59, 0.54, 0.59)$$

Therefore, the eigenvalues $\lambda_n$ reproduce quite accurately the experimental data of array#1. However, the model predicts that $c_3$ is almost equally distributed over the three

PCCs, while either in the experimental data or in the FDTD simulations the central cavity has very small electric field intensity. This discrepancy on the $T_3$ mode distribution likely suggests that the predictions of the analytical model are seminal but not fully correct, since the real system is more complex than three coupled oscillators. Moreover, the fact that in the K-configuration geometry, the coupling terms $g_K$ and $d_K$ have a strength comparable to the disorder induced energy detuning, is reflected in Fig. 2e), where different realization of the same K aligned array show a not unique spectral disposition of the three resonances. This large variety of behaviours is explained by adding in the K-matrix (5) the parameters $\Delta_l$ and $\Delta_c$ that account for the detuning of the lateral and central resonator, respectively. The mode spectral shift due to fabrication induced disorder introduces a net detuning between single PCCs. However, it can be compensated and slightly adjusted by post fabrication techniques able to locally modify the dielectric environment of the nanocavities, thus giving a net reduction of the detuning even if the fluctuations in the photonic holes are unchanged [34-36]. In this way it is possible to control on demand the photon hopping between adjacent nano-resonators. The system where the photon hopping rate is the same for every resonator is considered as the starting point for exploring quantum many-body phenomena with light [11,12]. Moreover, the post fabrication tuning methods also allow to introduce a uniform and controlled gradient in the photon hopping rate between different cavities, thus breaking the one-dimensional translational symmetry and pushing the system to a Bloch oscillations regime, which can be observed if the coherence times are greater than the period of the Bloch oscillations [37].

**CONCLUSION:**

In conclusion, we showed that different PCCs array configuration geometries lead to different results in terms of mode splitting and spatial distribution. In fact, in the case where both the nearest and next-nearest-neighbour coupling terms are relevant, the photon hopping between adjacent nano-resonators is strongly affected. These findings could open the way to exploit exotic configurations of coupled PCCs to tailor the mode spatial distribution or the group velocity in CROW of any length by engineering the dielectric properties of adjacent resonators in high-density optical circuits.

**ACKNOWLEDGMENTS:**

This work was supported by the FET project FP7 618025 CARTOON.

**FIGURES:**

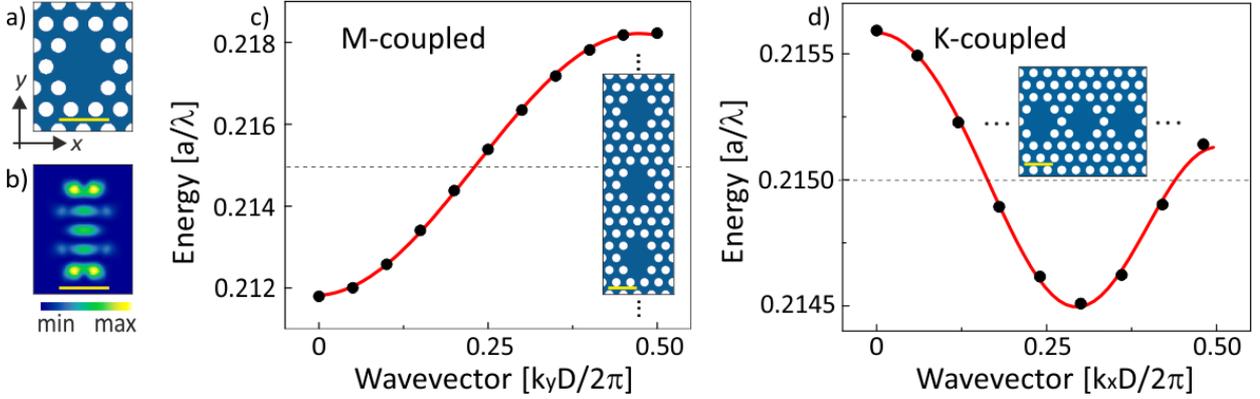

**Figure 1**: a)-b) Single PCC schematics and intensity distribution of the lower energy resonant mode, respectively. c)-d) Dispersion relation (black dots) of two-dimensional infinite array of aligned D2 nanocavities, calculated by plane wave expansion method and reported in the energy range relative to the fundamental mode of the single cavity, whose energy is highlighted by the horizontal dashed line. c) Calculation relative to the M-coupled infinite array (whose schematics is shown in the inset), reported as a function of the wavevector component parallel to the y-axis, in the irreducible Brillouin zone with $D = 3\sqrt{3}a$. The red curve provides the monotonous dispersion fit performed with Eq.(1). d) Calculation relative to the K-coupled infinite array (whose schematics is shown in the inset) reported as a function of the wavevector component parallel to the x-axis, in the irreducible Brillouin zone with $D = 3a$. The red curve provides the non-monotonous dispersion fit performed with Eq.(1). The scale bar in all the maps is 600 nm.

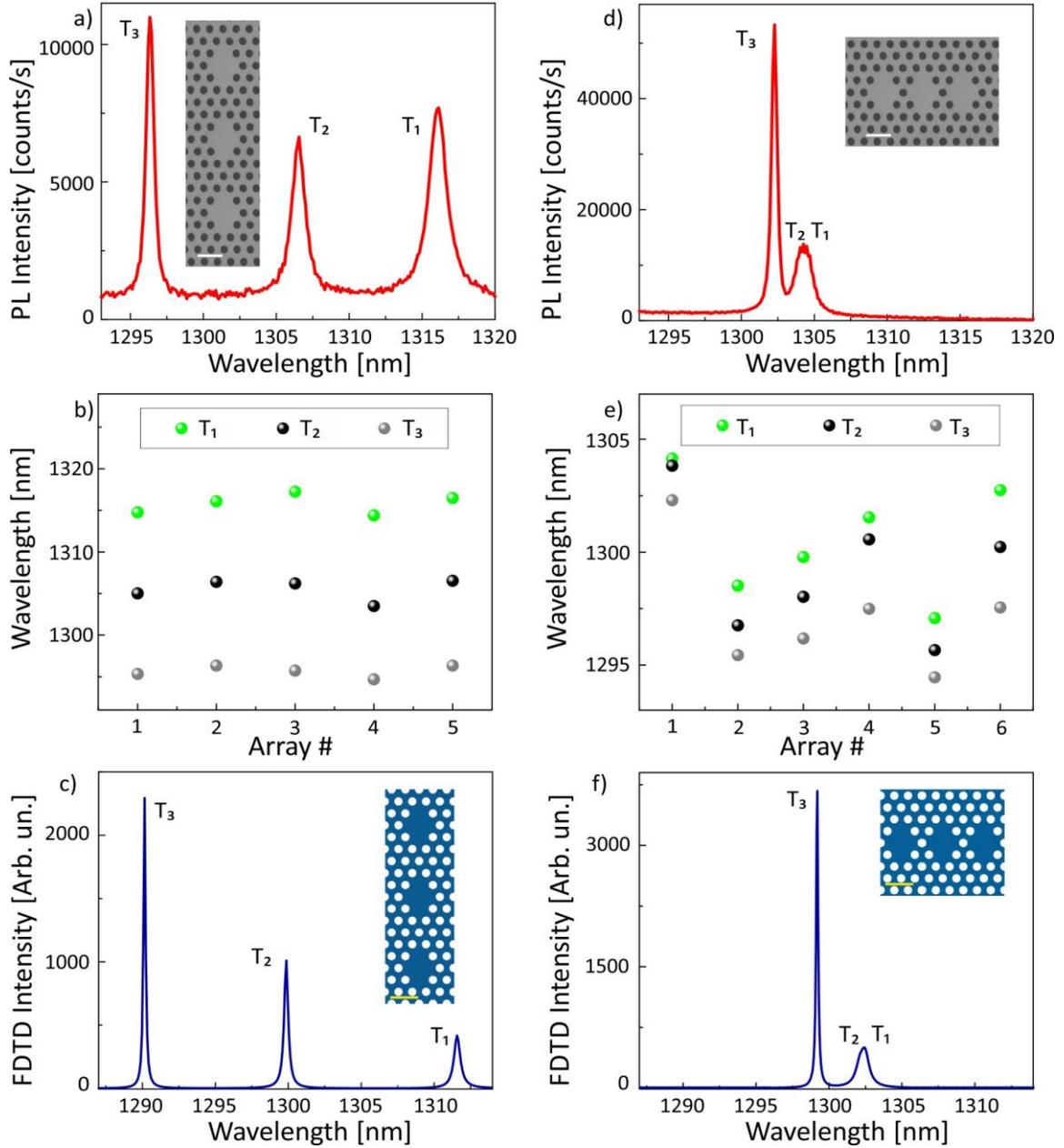

**Figure 2**: a) Experimental PL near-field spectrum averaged over the whole array structure of three M-coupled D2 nanocavities (see the SEM image in the inset). Three resonances (labelled $T_1, T_2, T_3$) are clearly observed. b) Experimental resonant modes wavelength values for five nominally identical M-coupled arrays, evaluated by fitting every peak with a Lorentzian function. The array#2 in b) corresponds to the case reported in a). c) Theoretical spectrum obtained by three-dimensional FDTD calculations, averaged over the M-coupled array structure reported in the inset. d)-f) Same analysis of a)-c) concerning the K-axis aligned array of three D2 nanocavities. In d) the modes $T_1$ and $T_2$ are almost degenerate as for the nominal design structure calculated by FDTD that is reported in f). e) Resonant modes wavelength values for six nominally identical K-coupled arrays. The array#1 in e) corresponds to the case shown in d). The scale bar in all the insets is 600 nm.

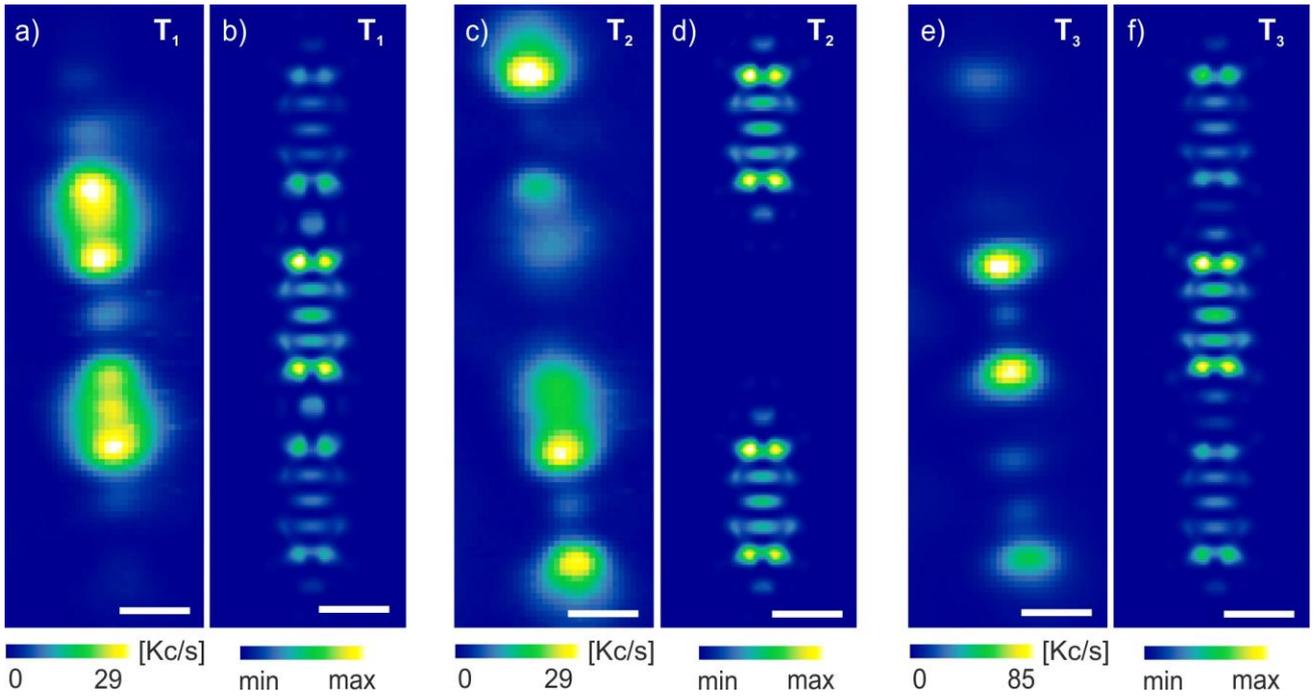

**Figure 3**: Experimental PL near-field map (in Kcounts/s) and FDTD calculated electric field intensity distribution, respectively, of the M-coupled array: a)-b) $T_1$ mode, c)-d) $T_2$ mode and e)-f) $T_3$ mode. The experimental maps are relative to the array#2 of Fig. 2b), whose spectrum is reported in Fig. 2a). The FDTD maps correspond to the peak wavelengths of the spectrum of Fig. 2c). The scale bar in all the figures is 600 nm.

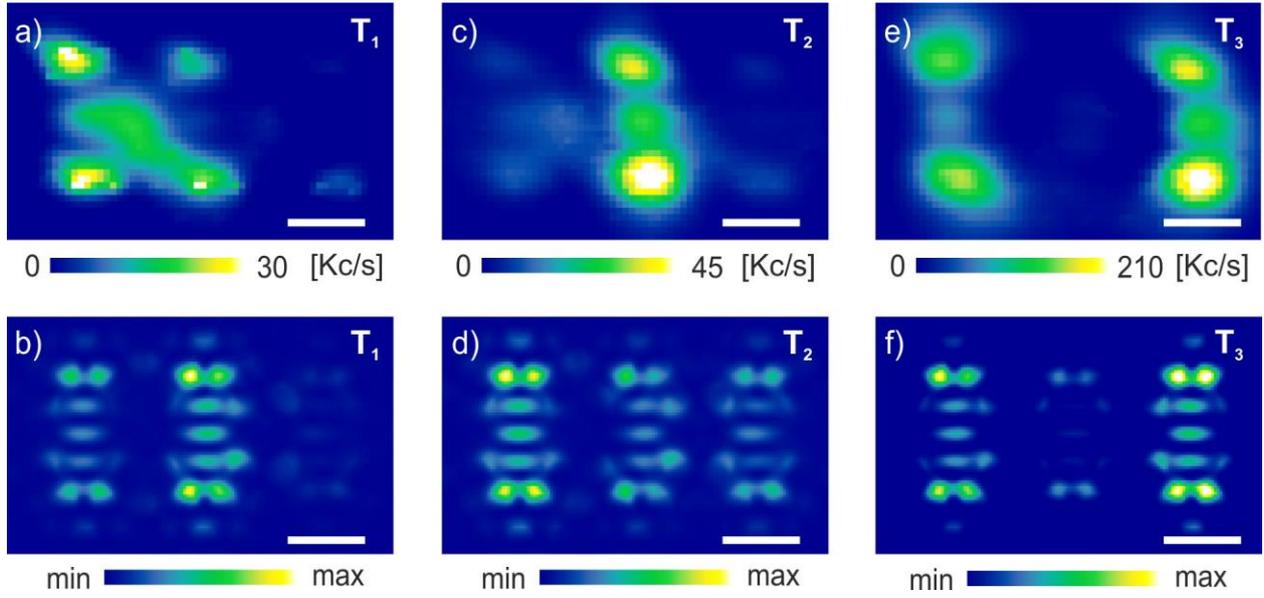

**Figure 4**: Experimental PL near-field map (in Kcounts/s) and FDTD calculated electric field intensity distribution, respectively, of the K-coupled array: a)-b) $T_1$ mode, c)-d) $T_2$ mode and e)-f) $T_3$ mode. The experimental maps are relative to the array#1 of Fig. 2e), whose spectrum is reported in Fig. 2d).The FDTD maps correspond to the peak wavelengths of the spectrum of Fig. 2f). The scale bar in all the figures is 600 nm.